\documentclass[prd,showpacs]{revtex4}

\usepackage{amssymb}
\usepackage{amsfonts}
\usepackage{amsmath}
\usepackage[dvips]{graphicx}

\begin{document}
  
\title{Acceleration of the Universe driven by the Casimir force}
 
 \author{Marek Szyd{\l}owski}
 \email{szydlo@oa.uj.edu.pl}
 \affiliation{Astronomical Observatory, Jagiellonian University,
 \\ Orla 171, 30-244 Krak{\'o}w, Poland}
 \affiliation{Complex Systems Research Centre, Jagiellonian University, 
 \\ Reymonta 4, 30-059 Krak{\'o}w, Poland}

 \author{W{\l}odzimierz God{\l}owski}
 \email{godlows@oa.uj.edu.pl}
 \affiliation{Astronomical Observatory, Jagiellonian University,
 \\ Orla 171, 30-244 Krak{\'o}w, Poland}

\begin{abstract}
 
We investigate an evolutional scenario of the FRW universe with the Casimir
energy scaling like $(-)(1+z)^4$. The Casimir effect is used to explain the 
vacuum energy differences (its value measured from astrophysics is so small
compared to value obtained from quantum field theory calculations).
The dynamics of the FRW model is represented in terms of a two-dimensional
dynamical system to show all evolutional paths of this model in the phase
space for all admissible initial conditions. We find also an exact solution 
for non flat evolutional paths of Universe driven by the Casimir effect. 
The main difference between the FRW model with the Casimir force and 
the $\Lambda$CDM model is that their generic solutions are a set of 
evolutional paths with a bounce solution and an initial singularity, 
respectively. The evolutional scenario are tested by using the SNIa
data, FRIIb radiogalaxies, baryon oscillation peak and CMB observation.
We compare the power of explanation of the model considered and the
$\Lambda$CDM model using the Bayesian information criterion and Bayesian
factor. Our investigation of the information criteria of model selection 
showed the preference of the $\Lambda$CDM model over the model considered. 
However the presence of negative like the radiation term can remove a tension 
between the theoretical and observed primordial ${}^4$He and D abundance.

\end{abstract}

\pacs{98.80.Jk, 04.20.-q}

\maketitle

\section {Introduction}
 
The recent astronomical observations like SNIa 
\cite{Riess:1998cb,Perlmutter:1998np}, WMAP \cite{Spergel:2003cb} indicate that 
the current Universe is in an accelerating phase of its expansion. While there 
are many different explanations of this phenomenon \cite{Copeland:2006wr} the
most straightforward one is that the Universe acceleration is due to the
presence of a dark energy component comprising more than $70\%$ of the energy 
of the Universe. Within this class of models the simplest candidate for dark 
energy is the phenomenological cosmological constant interpreted as vacuum 
energy \cite{Padmanabhan:2002ji}. Unfortunately a simple estimation from 
quantum field theory gives that vacuum energy is larger than what we observe by 
a factor of the order $10^{120}$. This discrepancy is called the cosmological
constant problem and we are looking for some physical process which works to
set vacuum energy precisely to zero. The Casimir force which is the 
manifestation of the quantum fluctuation can offer some mechanism to suppress 
the cosmological constant interpreted as vacuum energy to the value close to 
zero at the present epoch \cite{Ishak:2005xp,Antoniadis:2006wq,Bordag:2001qi,Ellingsen:2006qh}. 
In this context the Casimir effect \cite{Casimir:1948dh} seems to be relevant 
because it teaches us how vacuum energy contributes to the cosmological 
constant. Many other authors (see for example \cite{Milton:2001np}) argue that 
the Casimir effect is responsible for the vacuum energy differences between the 
value predicted by the quantum field theory and obtained from observations.
Recently Antoniadis et al. \cite{Antoniadis:2006wq} has suggested an important 
role of the Casimir effects in dark energy problem.
 
Ishak \cite{Ishak:2005xp} pointed out the relevance of experiments at the
interface of astrophysics and quantum field, focusing on the Casimir effects.
Nontrivial topology of the Universe \cite{Lachieze-Rey:1995kj} was the
motivations of these investigations in the cosmological context. Zeldovich
and Starobinsky \cite{Zeldovich:1984vk} investigated the simple closed FRW 
models equipped in a 3-torus topology. The Casimir energy has been also studied 
in the context of compactification of extra-dimensions as an effective 
mechanism of dimensional reduction 
\cite{Szydlowski:1987vr,Szydlowski:1988xf,Szydlowski:1988jy}.
It is worthy to mention that the Casimir type of contribution which arises from 
the tachyon condensation is also possible \cite{McInnes:2006uz}.
 
For example in a toroidal model with the compactification scale $L$, one 
typically obtains a Casimir contribution 
$\rho= \langle T^0_0 \rangle = -\frac{\alpha}{L^4}{a^4}$  
\cite{Mostepanenko:1997} (see also \cite{Muller:2004bh}) with the scale factor 
$a$ and the constant term $\alpha$ which depends on the nature and the number 
of matter fields. The analogous result was obtained by Wreszinski who used the 
``cosmic box'' idea of Harrison \cite{Wreszinski:2006dv} to explain why the 
cosmological constant assumes an absurdally small value of energy density. For 
this aim local theory of the Casimir effect was applied to the Universe as a 
whole. The case of massless scalar field with no potential was also 
investigated in the cosmological context by Hardeiro and Sampaio 
\cite{Herdeiro:2005zj}. The authors study backreaction of the metric on quantum 
effects in terms of fluid with Casimir energy 
$\rho_{\text{Cas}}=\alpha/a^4$, pressure $p_{\text{Cas}}=\alpha/3a^4$ and the 
sign of $\alpha$ for a massless scalar as a function of coupling $\xi$. The 
constant $\alpha$ changes the sign corresponds to a theory where the 
renormalized energy vanishes in the Einstein static universe. This critical 
value is $\xi_{\text{crit}} \simeq 0.05391$.
 
The massless conformally coupled scalar field as well as the electromagnetic
field and the massless Dirac field on the background of the Einstein static
universe (or equivalently quantum effects calculated in the adiabatic
approximations which can be justified by the fact that characteristic
time scale of quantum process is smaller than a scale of time evolution)
were considered in Refs.~\cite{Dowker:1976pr,Olver:1974,Streeruwitz:1975sv}.
In all cases the Casimir energy is of the form $\alpha/a^4$; when $\alpha$
is positive, than Casimir energy is an attractive force, while in an opposite
case is repulsive \cite{Zeldovich:1984vk}. In the latter case Zeldovich and 
Starobinski \cite{Zeldovich:1984vk} suggested that the scalar field could drive 
inflation in a flat universe with a nontrivial-toroidal topology. It is 
interesting that a dynamical effect of the Casimir force is dynamically
equivalent to effects of loop quantum gravity effects \cite{Mulryne:2005ef}.

The FRW model with the Casimir type force contains a term which scales like 
negative radiation $(-)(1+z)^4$. One should note that there are different 
interpretations of the presence of such a term: a cosmological model with 
global rotation, the Friedmann-Robertson-Walker (FRW) model in the 
Randall-Sundrum scenario with dark radiation, the FRW universe filled with a 
massless scalar field in a quantum regime or the FRW model in a semi-classical 
approximation of loop quantum gravity \cite{Godlowski:2006vf,Godlowski:2007gx}, 
however in the present paper the FRW universe filled with a massless scalar 
field in a quantum regime (Casimir effect) is of our special interest.
 
Our paper is organized as follows. In section II we provide some useful
dynamical background for further presentation. In section III we
discuss observational constraints on the parameters of model under
consideration. We complete the paper by Appendix in which we present an
exact solution in the bouncing cosmology with the Casimir effect.

\section{Dynamics of the FRW model with the Casimir dark radiation term}

In this section we investigate dynamical effects of the Casimir force which
phenomenologically can be modelled by the Casimir energy
$\rho_{\text{Cas}} \propto (-)a^{-4}$, where $a$ is the scale factor. It can be
shown that dynamical low temperature quantum effects of a scalar field can
be modelled in terms of such perfect fluid
\cite{Szydlowski:1987vr,Szydlowski:1988xf,Szydlowski:1988jy}. It is useful to 
investigate the FRW dynamics with the cosmological constant and the Casimir 
force using dynamical system methods. The main advantage of this tool is 
possibility of investigations of all evolutional path for all admissible 
initial conditions. The main feature of cosmological models with the Casimir 
force is the presence of the bounce instead of the standard initial 
singularity. It is worth mentioning that it is not possible by using a 
geometrical test (basing on null geodesics) to separate an individual Casimir 
component from all scaling in an analogous way, like $(1+z)^4$ 
\cite{Godlowski:2006vf,Godlowski:2007gx}.
 
We consider the FRW model filled by a perfect fluid with energy $\rho$ and
pressure $p$. Then the evolution of the scale factor $a$ can be described by
the simple acceleration equation
\begin{equation}
\frac{\ddot{a}}{a}=-\frac{1}{6}(\rho+3p),
\label{eq:b0}
\end{equation}
where $\rho$ and $p$ are function only the scale factor. It is convenient
to represent equation (\ref{eq:b0}) in term of dimensionless variable
$x=\frac{a}{a_0}$, where $a_0$ is the value of the scale factor at the
present epoch in the form analogous to the Newtonian equation of motion, i.e.
\begin{equation}
\ddot{x}=-\frac{\partial V}{\partial x}(x),
\label{eq:b1}
\end{equation}
where $V(x)$ is the potential function and overdot here denotes the
differentiation with respect to the rescaled cosmological time $t$ such that 
$t \to \tau \colon |H_0|dt=d\tau$, $H_0$ is the present value of the Hubble 
function.
 
The Newtonian equation of motion (\ref{eq:b1}) are equivalent to 
two-dimensional dynamical system of a Newtonian type
\begin{equation}
\dot{x}=y, \qquad \dot{y}=-\frac{\partial V}{\partial x}.
\label{eq:b1a}
\end{equation}
Therefore, admissible critical points
$y_0=0, \left. \frac{\partial V}{\partial x}\right|_{x_0}=0$
can be only saddles or centers.
 
Let us consider the Universe with the cosmological constant
($p_{\Lambda}=-\rho_{\Lambda}$, $\rho_{\Lambda}=\Lambda$) filled by
standard matter in which there are present the Casimir effect modelled as the
phenomological fluid scaling like $\rho_{\text{Cas}} \propto (-)a^4$. The 
assumed negative sign of energy $\rho_{\text{Cas}}$ denote that the Casimir 
force is attractive. The potential is
\begin{equation}
V(x) =-\frac{1}{2}( \Omega_{\Lambda,0} x^2 - \Omega_{k,0}+
\Omega_{\text{m},0} x^{-1} + \Omega_{\text{Cas},0} x^{-2}),
\label{eq:b6}
\end{equation}
where $\tau$ is new time parameter proportional to the original cosmological
time $t$ such that $t \to \tau \colon (H_0)dt=d\tau$; $x=a/a_0$ is a 
dimensionless scale factor expressed in the units of its present value $a_0$, 
$\Omega_{i,0}$ are the density parameters for matter ($i=\text{m}$), the 
Casimir contribution ($i=\text{Cas}$), the cosmological constant ($i=\Lambda$) 
and the curvature $k$ (the curvature index $k=0,\pm1$).
 
Equation (\ref{eq:b6}) admits the first integral
\begin{equation}
   \frac{(x')^2}{2}+V(x)=E \equiv 0
\label{eq:b7}
\end{equation}
which has a simple interpretation of conservation energy. The universe is 
presently ($x=1$) accelerating if $V(x)$ is a decreasing function of its
argument $x \colon \left(\frac{dV}{dx}\right)_{x=1}<0$.
 
\begin{figure}
\begin{center}
\includegraphics[width=0.9\textwidth]{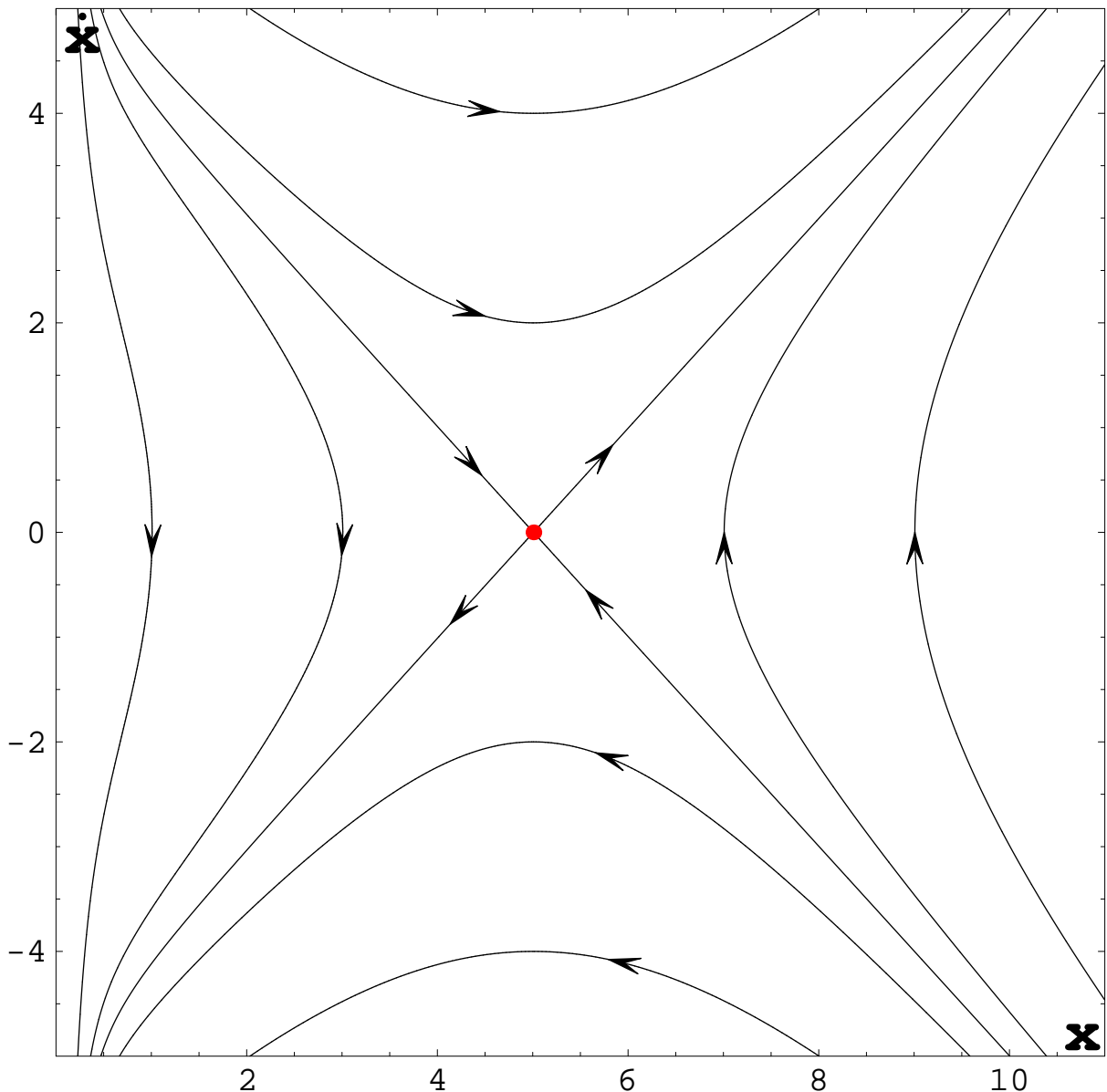}
\end{center}
\caption{Phase portrait for the FRW model with $\alpha>0$ (usual radiation)
and the cosmological constant $\Lambda$ (positive). There are three different
types of evolution oscillating, loitering and bouncing. The trajectory $k=0$
of the flat model separates all the evolutional paths of closed and open 
models.}
\label{fig:0a}
\end{figure}
 
\begin{figure}
\begin{center}
\includegraphics[width=0.9\textwidth]{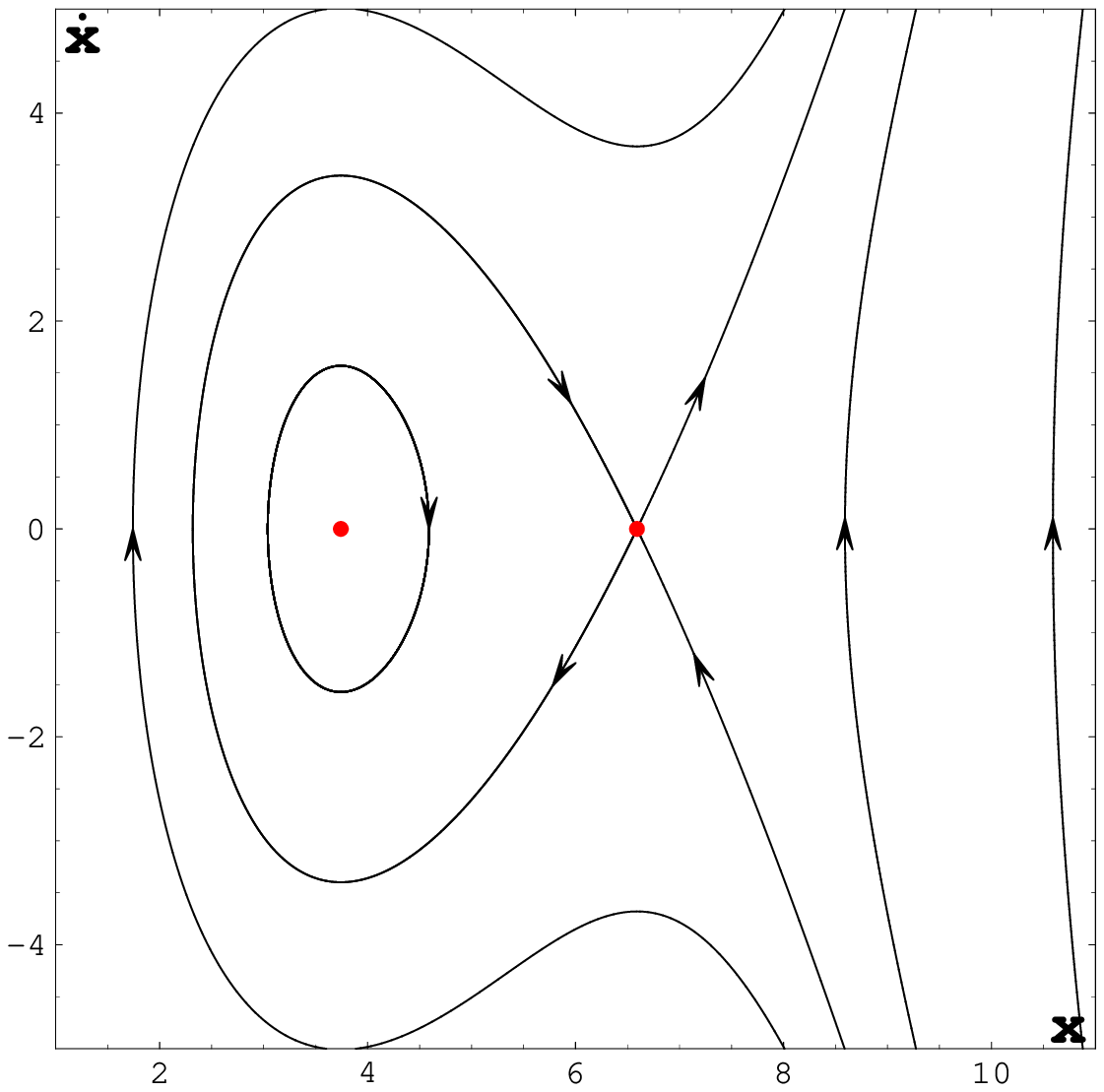}
\end{center}
\caption{Phase portrait for the FRW model with the repulsive Casimir force
($\alpha<0$). All models are undergoing a bounce instead of an initial
singularity replaced by the critical point $(x,\dot{x})=(0,\infty)$ like in
Fig.~1. In the phase portrait there are present two types of static critical
points center and saddle point like in Fig.~1.
Note that bounce appears to be possible in open flat and closed cosmologies.
The phase portrait is structurally unstable in contrast to the $\Lambda$CDM
model represented in Fig.~1.}
\label{fig:0b}
\end{figure}
 
From equation (\ref{eq:b7}) we obtain at the present epoch the constraint
$V(x=1)=-\frac{1}{2}$ or
\begin{equation}
   \sum_{i} \Omega_{i,0}=1.
\label{eq:b8}
\end{equation}
 
From equation (\ref{eq:b6}) we obtain that negative values of $\alpha$
as well as the positive cosmological constant give rise to acceleration
of the Universe. Because of constraint (\ref{eq:b7}) the motion of the
system takes place in a domain $\mathcal{D}_0={x \colon V(x)\le 0)}$ of the
configuration space.
 
If we shift the curvature term $-\frac{1}{2}\Omega_{k,0}$ on the right hand side
of (\ref{eq:b7}) then system (\ref{eq:b6}) can be considered on the constant
energy level $E=-\frac{1}{2}\Omega_{k,0}$ parametrized by the curvature
constant. The phase space $(x,x')$ of all trajectories for all admissible
initial conditions is divided by trajectory of the flat models $\Omega_{k,0}=0$.
Closed models ($\Omega_{k,0}<0$) are situated inside this characteristic
curve and open models ($\Omega_{k,0}>0$) outside.

The phase portraits together with the potential function (\ref{eq:b6}) are
shown in Fig. \ref{fig:0a} and Fig. \ref{fig:0b} for $\alpha<0$
and $\alpha>0$, respectively. From the phase portrait in Fig. \ref{fig:0b}
(for $\alpha<0$, $\rho_{\text{Cas}}=\alpha a^{-4}$) we can observe that the
initial singularity characteristic for the case of $\alpha>0$ is replaced
by a bounce. To have a bounce, there must be some time at which the size
of the universe (or the scale factor) assumes a minimum: $\dot{a}(t_0)=0$.
Moreover $\ddot{a}(t_0) \ge 0$ is required. The sufficient condition for the
origin of the FRW universe from a bounce was formulated by Molina-Paris and
Visser \cite{Molina-Paris:1998tx}. From the point of view of the structural stability
notion there is the main difference between phase portraits in Fig.~\ref{fig:0a}
and Fig.~\ref{fig:0b}. The phase portrait for the $\Lambda$CDM model is
structurally stable. This means that small perturbation of right-hand sides 
does not disturb the phase portrait. In contrast to Fig.~\ref{fig:0a} the phase
portrait in Fig.~\ref{fig:0b}, is structurally unstable due to presence of
centre on $x$-axis. Following the Peixoto theorem the $\Lambda$CDM dynamical
system is generic in the class of all planar dynamical systems and bouncing
cosmological models are exceptional because do not form such open dense
subsets. The type of a critical point is determined by eigenvalues of the
linearization matrix calculated at the critical point. This equation
assumes the form $\lambda^2 + \left. \frac{\partial^2V}{\partial x^2}\right|_{x=x_0}=0$.
Therefore the bounce representing by a centre is present on the phase portrait
if $V_{xx}(x_0)>0$, where $V_x(x_0)=0$, $x_0$ is the critical point
geometrically it means that $V(x)$ function has minimum in $x_0$.
 
Note that all models on the phase portraits behaves reflectional symmetry
$x \to (-)x$, $x' \to (-)x'$. During the bounce the SEC (strong energy 
condition) violation is necessary.
 
For present acceleration of the universe it is required
\begin{equation}
 -\Omega_{\text{m},0}-2\Omega_{\text{Cas},0}+\Omega_{\Lambda,0}>0.
\label{eq:b9}
\end{equation}
Therefore if $\Lambda=0$ the Casimir force must be sufficiently large
\begin{equation}
 (|\Omega_{\text{Cas},0}|+\Omega_{\Lambda,0})>\frac{\Omega_{\text{m},0}}{2}
\label{eq:b10}
\end{equation}

The bouncing cosmology is following of Barrow's reincarnation of ancient
fascination of cyclicity and realization myth of the "eternal return"
\cite{Barrow:2004ad}. While the classical bounces are generated by fields
endowed with negative energy, the quantum gravity effects in semi-classical
approximation invoked to justify the bounce \cite{Singh:2003au} generated by
some quantum corrections terms. Also possibility that a contracting
braneworld model experiences a bounce instead ever reaching a singularity
is addressed by Shtanov and Sahni 
\cite{Shtanov:2000vr,Shtanov:2002ek,Sahni:2002dx,Shtanov:2002mb}.

Recently bouncing cosmology has been analyzed in details in 
\cite{Szydlowski:2005qb}. It was also demonstrated that the probability of a 
bounce is close to unity for models of $f(R)$ nonlinear gravity 
\cite{Borowiec:2006hk,Borowiec:2006qr,Carloni:2005ii}.

\begin{figure}
\begin{center}
\includegraphics[width=0.9\textwidth]{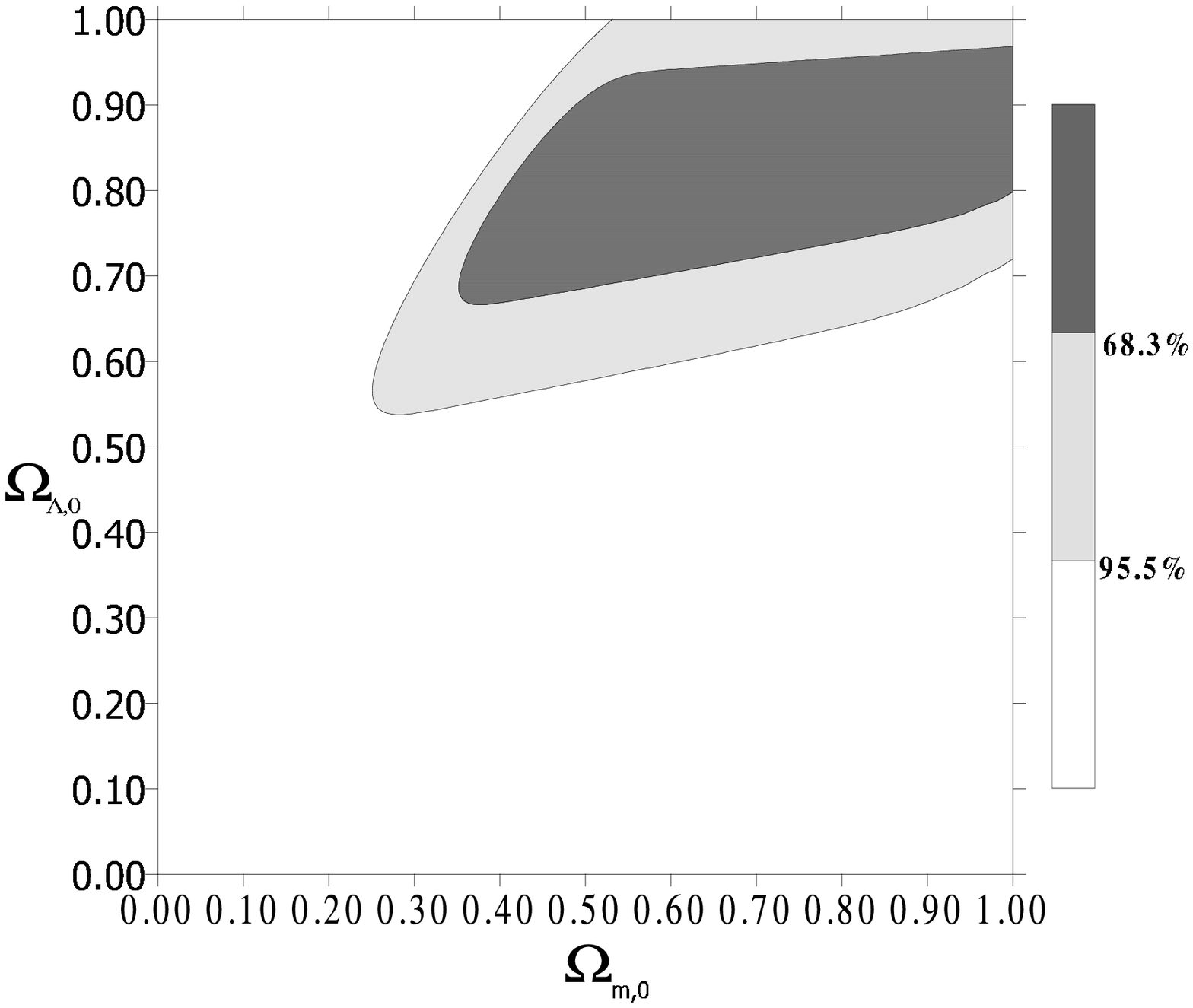}
\end{center}
\caption{The $68.3\%$ and $95.4\%$ confidence levels obtained from combined
analysis SN+RG) on the ($\Omega_{\text{m},0},\Omega_{\Lambda,0}$) plane.}
\label{fig:1}
\end{figure}

\begin{figure}
\begin{center}
\includegraphics[width=0.9\textwidth]{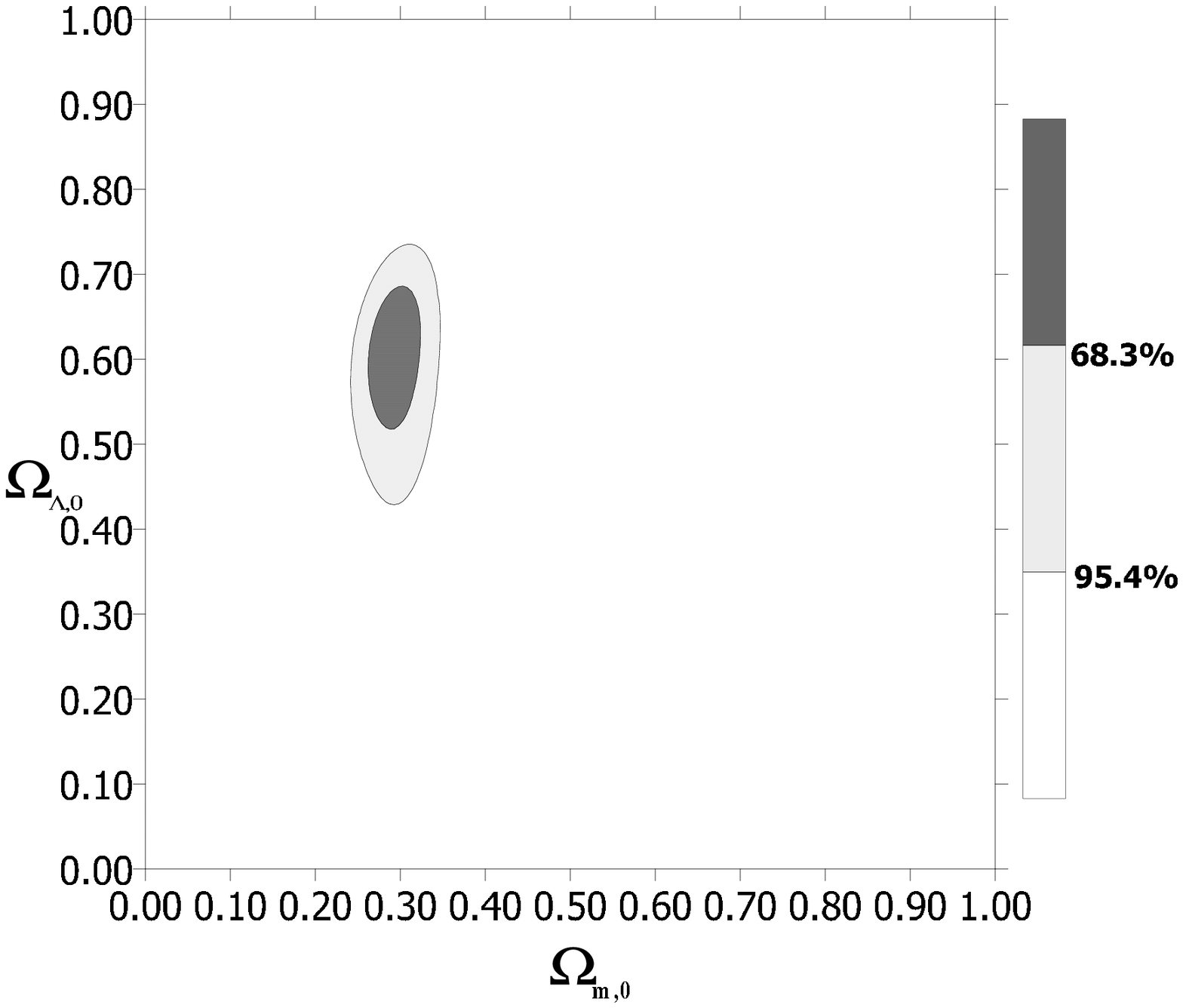}
\end{center}
\caption{The $68.3\%$ and $95.4\%$ confidence levels obtained from combined
analysis SN+RG+BOP) on the ($\Omega_{\text{m},0},\Omega_{\Lambda,0}$) plane.}
\label{fig:2}
\end{figure}

\begin{figure}
\begin{center}
\includegraphics[width=0.9\textwidth]{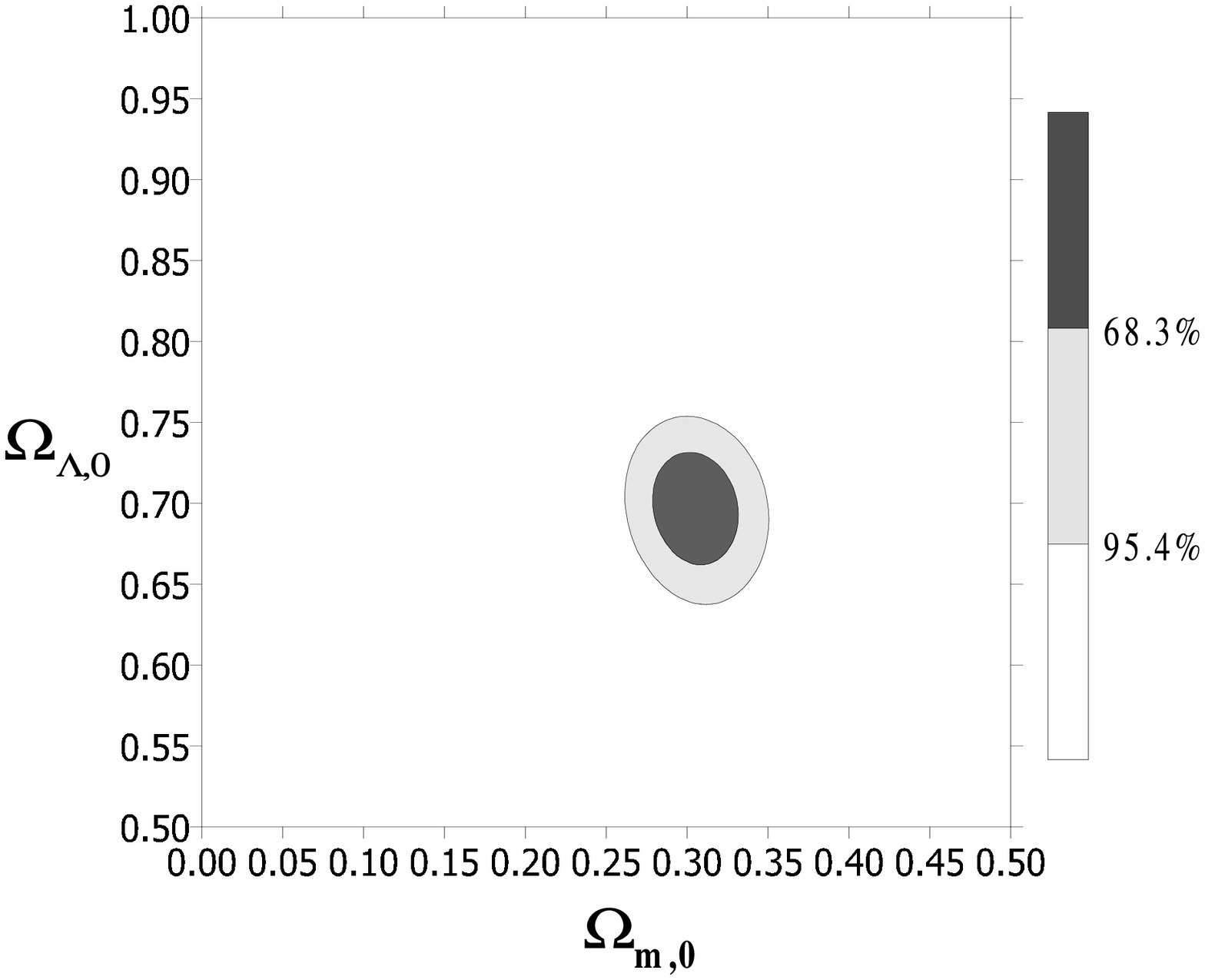}
\end{center}
\caption{The $68.3\%$ and $95.4\%$ confidence levels obtained from combined
analysis SN+RG+BOP+CMB) on the ($\Omega_{\text{m},0},\Omega_{\Lambda,0}$) plane.}
\label{fig:3}
\end{figure}

\section{Toward testing bouncing cosmology caused by $(-)(1+z)^4$ term}

It is very interesting to test cosmological models against observations.
One of the most popular test used the SNIa data. This test is based
on the luminosity distance $d_L$ of the supernovae Ia as a function of
redshift \cite{Riess:1998cb}. The observations of the type Ia distant supernova
suggest that the present Universe is accelerating
\cite{Riess:1998cb,Perlmutter:1998np,Riess:2004nr,Riess:2006fw}. Every year new 
SNIa enlarge the available data by more distant objects and lower systematics
errors. Our work is based on two samples: Riess et al. \cite{Riess:2004nr}
``Gold'' sample of 157 SNIa and Astier et al. \cite{Astier:2005qq} sample of
supernovae, based on 71 high redshifted SNIa discovered during the first
year of the 5-year Supernovae Legacy Survey. We used this sample because
we would like to easily compare the result obtained in present paper with
that obtained, with using of different method, in our previous paper
\cite{Godlowski:2006vf}.
 
One should note that for the distant SNIa, we directly observe their apparent
magnitude $m$ and redshift $z$.  The absolute magnitude $\mathcal{M}$ of the
supernovae is related to its absolute luminosity $L$. We have the following
relation between distance modulus $\mu$, the luminosity distance $d_L$,
the observed magnitude $m$ and the absolute
magnitude $M$:
\begin{equation}
\label{eq:11}
\mu \equiv  m - M = 5\log_{10}d_{L} + 25=5\log_{10}D_{L} + \mathcal{M}
\end{equation}
where $D_{L}=H_{0}d_{L}$ and $\mathcal{M} = - 5\log_{10}H_{0} + 25$.
The luminosity distance of a supernova is the function of cosmological
parameters and redshift
\begin{equation}
\label{eq:12}
d_L(z) =  (1+z) \frac{c}{H_0} \frac{1}{\sqrt{|\Omega_{k,0}|}} \mathcal{F}
\left( H_0 \sqrt{|\Omega_{k,0}|} \int_0^z \frac{d z'}{H(z')} \right)
\end{equation}
where
\begin{equation}
\label{eq:12a}
\left(\frac{H}{H_0}\right)^2= \Omega_{\text{m},0}(1+z)^{3}+\Omega_{k,0}(1+z)^{4}+
\Omega_{\text{r},0}(1+z)^{4}+\Omega_{\text{dr},0}(1+z)^{4}+\Omega_{\Lambda,0},
\end{equation}
$\Omega_{k,0} = - \frac{k}{H_0^2}$ and
$\mathcal{F} (x) \equiv (\sinh (x), x,\sin (x))$ for $k<0, k=0, k>0$,
respectively. We assumed $\Omega_{\text{r},0} = \Omega_{\gamma,0} + \Omega_{\nu,0}
= 2.48 h^{-2} \times 10^{-5} + 1.7 h^{-2} \times 10^{-5}\simeq 0.0001$
\cite{Vishwakarma:2002ek}.
 
Substituting (\ref{eq:12}) back into equations (\ref{eq:11})
provides us with an effective tool (the Hubble diagram) to test cosmological
models and to constrain their parameters. Assuming that supernovae
measurements come with uncorrelated Gaussian errors, one can determine the
likelihood function $\mathcal{L}$   from  chi-square
statistic $\mathcal{L}\propto \exp(-\chi^{2}/2)$, where
\begin{equation}
\label{eq:10}
\chi^{2}=\sum_{i}\frac{(\mu_{i}^{\mathrm{theor}}-\mu_{i}^{\mathrm{obs}})^{2}}
{\sigma_{i}^{2}} \ .
\end{equation}
The probability density function (PDF) of cosmological parameters
\cite{Riess:1998cb} can be derived from Bayes' theorem. Therefore, one can
estimate model parameters by using a minimization procedure. It is based on
the likelihood function as well as on the best fit method  minimizing $\chi^2$.
Constraints for the cosmological parameters, can be obtain by minimizing
the following likelihood function $\mathcal{L}\propto \exp(-\chi^{2}/2)$.

Daly and Djorgovski \cite{Daly:2003iy} (see also
\cite{Zhu:2004ij,Puetzfeld:2004sw,Godlowski:2006vf,Godlowski:2007gx})
suggested to include in the analysis not only supernovae but also radio
galaxies. In such a case, it is useful to use the coordinate distance
defined as \cite{Weinberg:1989}
\begin{equation}
\label{eq:12b}
y(z)=\frac{H_0 d_L(z)}{c(1+z)}.
\end{equation}
In such a case we can determinate likelihood function
$\mathcal{L}\propto \exp(-\chi^{2}/2)$, where
\begin{equation}
\label{eq:13}
\chi^{2}=
\sum_{i}\left(\frac{y_{i}^{\text{obs}}-y_{i}^{\text{th}}}{\sigma_{i}(y_{i})}\right)^{2}
\end{equation}

To analyse SNIa data and FRIIb data in the unique way it is useful analises
for both sets of data coordinate distance $y(z)$. Daly and Djorgovski
\cite{Daly:2004gf} compiled a sample comprising the data on $y(z)$ for
157 SNIa in the Riess et al. \cite{Riess:2004nr} Gold dataset and 20 FRIIb
radio galaxies. In our data sets we also include 115 SNIa compiled
by Astier et~al. \cite{Astier:2005qq}. In our previous paper \cite{Godlowski:2006vf}
there are more details on the computation of $y(z)$ and $\sigma_{i}(y_{i})$ for 
these samples.
 
In Fig.~\ref{fig:1} we present the $68.3\%$ and $95.4\%$ confidence levels 
obtained from combined analysis SNIa and FRIIb on the plane 
($\Omega_{\text{m},0},\Omega_{\Lambda,0}$).

We also include to our analysis the baryon oscillation peaks (BOP)
detected in the Sloan Digital Sky Survey (SDSS) luminous red galaxies
\cite{Eisenstein:2005su}. They found that the value $A$ is
\begin{equation}
\label{eq:16}
A \equiv \frac{\sqrt{\Omega_{\text{m},0}}}{E(z_1)^{\frac{1}{3}}}
\left(\frac{1}{z_1\sqrt{|\Omega_{k,0}|}}
\mathcal{F} \left( \sqrt{|\Omega_{k,0}|} \int_0^{z_1} \frac{d z}{E(z)} \right)
\right)^{\frac{2}{3}} =0.469 \pm 0.017
\end{equation}
where $E(z) \equiv H(z)/H_0$ and $z_1=0.35$.
The quoted uncertainty corresponds to one standard deviation, where a Gaussian
probability distribution has been assumed. These constraints could also be
used for fitting cosmological parameters \cite{Astier:2005qq,Fairbairn:2005ue}.
In such a case we determinate the likelihood function
$\mathcal{L}\propto \exp \left[-\left(\frac{A^{\text{mod}}-0.469}{0.017}\right)^{2}/2\right]$.

Another constraint which we also include in our analysis is the
so called the (CMBR) ``shift parameter'' \cite{Wang:2004py}
\begin{equation}
\label{eq:17}
R \equiv \sqrt{\Omega_{\text{m},0}} \, y(z_{\text{lss}})=
\sqrt{\frac{\Omega_{\text{m},0}}{|\Omega_{k,0}|}}
\mathcal{F} \left(\sqrt{|\Omega_{k,0}|} 
\int_0^{z_{\text{lss}}} \frac{d z}{E(z)}\right)
=1.716 \pm 0.062
\nonumber
\end{equation}
which leads to the likelihood function 
$\mathcal{L}\propto \exp(-\left(\frac{R^{\text{mod}}-1.716}{0.062}\right)^{2}/2)$.

In Fig.~\ref{fig:2} and Fig.~\ref{fig:3} we present confidence levels obtained
from the analysis of SNIa, FRIIb and BOP or BOP and CMB shift, respectively.
One should note that when we are interested in constraining a particular model
parameter, the likelihood function marginalized over the remaining parameters
of the model should be considered \cite{Cardone:2005ut}.

From our combined analysis (SN+RG+SDSS+CMBR) we obtain as the best fit
a flat (or nearly flat universe) with $\Omega_{\text{m},0} \simeq 0.3$,  and
$\Omega_{\Lambda,0} \simeq 0.7$. For the dark radiation term, we obtain the
stringent bound
$\Omega_{\text{R},0}=\Omega_{\text{r},0}-|\Omega_{\text{dr},0}| > -0.00025$
at the 95\% confidence level where $\Omega_{\text{dr},0}$ is negative.
It leads for the limit on dark radiation $|\Omega_{\text{dr},0}| < 0.00035$.
This results mean that null value of dark radiation term is preferred
($\Omega_{\text{dr},0}=0$), however the small negative contribution of dark
radiation is also available. Our results shows that in the present epoch
contribution of the dark radiation, if it exist, is small
and gives only small corrections to the $\Lambda$CDM model in the low redshift.

Our result is in agreement with the result of our previous paper 
\cite{Godlowski:2006vf} where in the combined analysis based in a 
pseudo-$\chi^2$ merit function \cite{Cardone:2005ut}: we obtain as the best fit
a flat universe with $\Omega_{\text{m},0}=0.3$, $\Omega_{\text{dr},0}=0$ and
$\Omega_{\Lambda,0}=0.7$. For the dark radiation term, we obtain the stringent
bound $|\Omega_{\text{dr},0}| < 0.00035$ at the 95\% confidence level
($|\Omega_{\text{dr},0}| <0.00026$ at the $68.3\%$ confidence level).

Please also note that if
$\Omega_{\text{R},0}=\Omega_{\text{r},0}-|\Omega_{\text{dr},0}|<0$,
then we obtain a bouncing scenario 
\cite{Molina-Paris:1998tx,Tippett:2004xj,Szydlowski:2005qb}
instead of a big bang.
For $\Omega_{\text{m},0}=0.3$, $\Omega_{\text{dr},0}=-0.00035$ and $h=0.65$
bounces ($H^2=0$) appear for $z \simeq 1200$. In this case, the BBN epoch
never occurs and all BBN predictions would be lost. 

We use the Akaike information criterion (AIC) \cite{Akaike:1974}, the
Bayesian information criterion (BIC) \cite{Schwarz:1978} as well as the
Bayes factor to select model parameters providing the preferred fit to data. 
The information criteria
put a threshold which must be exceeded in order to assert an additional
parameter to be important in explanation of a phenomenon. The discussion how
high this threshold should be caused the appearing of many different criteria.
The Akaike and Bayesian information criteria (AIC and BIC) (for review see
\cite{Burnham:2002}) are most popular and used in everyday statistical practices.
The usefulness of using the information criteria of model selection was
recently demonstrated by Liddle \cite{Liddle:2004nh} and Parkinson et al.
\cite{Parkinson:2004yx}. The problem of classification of the cosmological
models on the light of the information criteria on the base of the
astronomical data was discussed in our previous papers
\cite{Godlowski:2005tw,Szydlowski:2005xv,Szydlowski:2005bx,Szydlowski:2006gp,Szydlowski:2006ay,Kurek:2007tb}.

The AIC is defined in the following way \cite{Akaike:1974}
\begin{equation} \label{eq:111}
\text{AIC} = - 2\ln{\mathcal{L}} + 2d
\end{equation}
where $\mathcal{L}$ is the maximum likelihood and $d$ is a number of the
free model parameters. The best model with a parameter set providing the
preferred  fit to the data is that minimizes the AIC.
The BIC introduced by Schwarz \cite{Schwarz:1978} is defined as
\begin{equation} \label{eq:112}
\text{BIC} = - 2\ln{\mathcal{L}} + d\ln{N}
\end{equation}
where $N$ is the number of data points used in the fit. Comparing these
criteria, one should note that the AIC tends to favor models
with large number of parameters unlike the BIC, because the BIC
penalizes additional parameters more strongly.
Of course only the relative value between the BIC of different
models has statistical significance. The difference of $2$ is treated as
a positive evidence (and $6$ as a strong evidence) against the model with
the larger value of the BIC \cite{Jeffreys:1961,Mukherjee:1998wp}.
If we do not find any positive evidence from the information criteria the
models are treated as a identical and eventually additional parameters are
treated as not significant.

In the Bayesian framework quality of the models can be compared with help
of evidence \cite{Jeffreys:1961,Mukherjee:2005tr}. We can define the 
a'posterior odds for two models $M_i$ and $M_j$ -- the Bayes factor $B_{ij}$ 
\cite{Kass:1995}. If we do not favor any model it reduces to the evidence 
ratio. Schwarz \cite{Schwarz:1978} showed that for observations coming from 
a linear exponential family distribution the asymptotic approximation 
$N \to \infty$ the logarithm of evidence is given by
\begin{equation} \label{eq:113}
\ln{E}=\ln{\mathcal{L}} - \frac{d}{2}\ln{N} +O(1).
\end{equation}
It is easy to show that in this case we have the simple relation between
the Bayes factor and the BIC
\begin{equation} \label{eq:114}
2\ln{B_{ij}} = -(\text{BIC}_i-\text{BIC}_j)
\end{equation}

If $B_{ij}$ is greater than 3 it is called positive evidence in favor of $M_i$
model, while  $B_{ij}>20$ give strong and  $B_{ij}>150$ very strong evidence
in favor of $M_i$ model \cite{Szydlowski:2006ay}.

Our results are presented in Table~\ref{tab:1}. In our investigations we 
consider the likelihood function $\mathcal{L}\propto \exp(-\chi^{2}/2)$
where a pseudo-$\chi^2$ merit function is used 
\cite{Cardone:2005ut,Godlowski:2006vf,Godlowski:2007gx}:
\begin{equation}
\label{eq:18}
\chi^{2}=\chi_{\text{SN+RG}}^{2}+\chi_{\text{SDSS}}^{2} +\chi_{\text{CMBR}}^{2}=
\nonumber
\end{equation}
\begin{equation}
\sum_{i}\left(\frac{y_{i}^{\text{obs}}-y_{i}^{\text{th}}}{\sigma_{i}(y_{i})}\right)^{2}+
\left(\frac{A^{\text{mod}}-0.469}{0.017}\right)^{2}+
\left(\frac{R^{\text{mod}}-1.716}{0.062}\right)^{2},
\end{equation}
where $A^{\text{mod}}$ and $R^{\text{mod}}$ denote the values of $A$ and $R$
obtained for a particular set of the model parameter.
 
Our investigations of the information criteria show that the bouncing $\Lambda$ 
cold dark matter model (B$\Lambda$CDM) with dark energy does not increase the 
fit significantly. It confirms our conclusion that the dark energy term, if it 
exists, is small in the present epoch. The Bayes factor also favors the 
$\Lambda$CDM model over the B$\Lambda$CDM model with dark energy. Its supports 
results obtained with help of the AIC and BIC. One should note that because of 
a non-Gaussian distribution of the a'posteriori PDF function for $\Omega_{R,0}$ 
the above results obtained with help of the Bayes factor should be treated with 
caution as only an additional support for results obtained with help of the AIC 
and BIC.

\section{Conclusion}

In this paper we analyzed the observational constraints on the $(1+z)^4$-type
contribution in the Friedmann equation. The analysis of SNIa data as well as
both SNIa and FRIIb radio galaxies with the constraints coming from baryon
oscillation peaks and CMBR ''shift parameter`` shows that influence of a
negative term  $-(1+z)^4$  is very weak in the present epoch of the Universe.
 
The main aim of this paper is investigation of evolutionary paths of the FRW
cosmological models with the cosmological constant $\Lambda$ and the Casimir
energy scaling like $(-)(1+z)^4$. We demonstrate that a bounce is a generic 
feature of this model. In this scenario an initial singularity is replaced by 
a bounce. We characterize a full class of all evolutionary paths by using 
dynamic system methods and we find exact solution. We pointed out the main
difference between the $\Lambda$CDM model and the $B\Lambda$CDM model from the
view of point structural stability. The two-dimensional case of dynamical
system is distinguished by the fact that the Peixoto theorem gave a complete
characterization of the structurally stable systems on any compact space, assert
that they form open and dense subsets in the space of all dynamical systems on
the plane. We conclude that while the $\Lambda$CDM models are structurally
stable, therefore typical, the B$\Lambda$CDM models are structurally unstable, 
therefore exceptional. Of course structurally unstable models may  model a 
realistic physical situation however they require a fine tuning of the model 
itself (not fine tuning of initial conditions).
 
We show that there are several interpretations of the $(1+z)^4$-type
contribution and we discussed different proposals for the presence of such a 
term. Unfortunately, it is not possible, with present kinematic astronomical 
tests, to determine the energy densities of individual components scales like 
radiation. However we show that some stringent bounds on the value of this 
total contribution can be given. The combined analysis of SNIa data and FRIIb 
radio galaxies, baryon oscillation peaks and CMBR ''shift parameter`` give 
rise to the concordance universe model which is almost flat with 
$\Omega_{\text{m},0} \simeq 0.3$. From the above-mentioned combined analysis, 
we obtain an constraint for the negative term which scales like total radiation
$\Omega_{\text{R},0} > -0.00025$ which leads to bounds on the dark radiation 
term $\Omega_{\text{dr},0} > -0.00035$ at the $95\%$ confidence level. This is 
in agreement with result of our previous paper \cite{Godlowski:2006vf} obtained 
with help of a pseudo-$\chi^2$ merit function.
 
The investigations of the information criteria show the B$\Lambda$CDM model 
with dark energy does not increase the fit significantly. It confirms our 
conclusion that the dark energy term, if it exists, is small in the present 
epoch.
 
In this paper we have been especially studied the advance of an initial
singularity using back reaction gravity quantum effect at low temperatures
(the Casimir effect). The Casimir force arising from the quantum effect of
massless scalar field give rise to a $(-)(1+z)^4$ correction whose effect
depends upon the geometry and nontrivial topology of the space. Typically
this type of correction is thought to be important at the late time of
evolution of the universe. We have shown that the Casimir effect could
generically remove the initial singularity which would be replaced by the
bounce.
 
Our analysis of back reaction on quantum effect clearly reflects an
important role played by vacuum energy. The basic problem which helps us to
understand the dynamics of accelerating universe is how the cosmological
constant (geometrical) contributes to the vacuum energy.

\section*{Acknowledgements}
M. S. was supported by the Marie Curie Actions Transfer of Knowledge 
project COCOS (contract MTKD-CT-2004-517186). Authors are grateful 
T. Stachowiak for comments and fruitful discussion. The authors also thank 
Dr. A. G. Riess, Dr. P. Astier and Dr. R. Daly for the detailed explanation 
of their data samples.

\begin{table}
\caption{The values of AIC and BIC and Bayes factor
$B_{12}$ for $\Lambda$CDM model and bouncing cosmology model (with dark
radiation). The upper section of the table represents the constraint
$\Omega_{k,0}=0$ (flat model).}
\begin{tabular}{c|cc|cc|c}
\hline \hline
\multicolumn{1}{c}{}&
\multicolumn{2}{c}{$\Lambda$CDM}&
\multicolumn{2}{c}{B$\Lambda$CDM}&
\multicolumn{1}{c}{} \\
\hline \hline
sample & AIC & BIC & AIC & BIC & $B_{12}$ \\
\hline
SN              & 299.5& 303.1& 301.5 & 308.7 & 16.44  \\
SN+RG           & 322.4& 326.1& 324.4 & 331.8 & 17.29  \\
SN+RG+SDSS      & 324.4& 328.1& 326.4 & 333.8 & 17.29  \\
SN+RG+SDSS+CMBR & 324.5& 328.2& 326.5 & 333.9 & 17.29  \\
\hline
SN              & 300.0& 307.2& 302.0 & 312.8 & 16.44  \\
SN+RG           & 323.5& 330.9& 325.5 & 336.5 & 17.29  \\
SN+RG+SDSS      & 325.1& 332.5& 327.1 & 338.1 & 17.29  \\
SN+RG+SDSS+CMBR & 326.5& 333.9& 328.5 & 339.5 & 17.29 \\
\hline
\end{tabular}
\label{tab:1}
\end{table}

\appendix

\section{Basic models with the Casimir force.}

The Friedmann equation we are dealing with can be reduced so that the
right-hand side is a 4th degree polynomial. Such equations have been
completely analyzed in \cite{Dabrowski:2004hx}, so we merely point out
which cases of the general classification (by which we will mean the cited
paper's content) apply and what additional features arise here.

The equation reads:
\begin{equation}
    \left(\frac{d y}{d u}\right)^2 = \Omega_{\Lambda,0} y^2 - \Omega_{k,0}
    + \Omega_{m,0} y^{-1} + \Omega_{r,0} y^{-2} - \Omega_{\omega} y^{-2}.
\end{equation}

Analogously to the $\Omega_{R,0}$ in section III we introduce
$\Omega_{R} = \Omega_{r,0} - \Omega_{\omega}$, and change the variables to:
\begin{equation*}
    y = x, \quad du = x d\tau,
\end{equation*}
the main equation becomes
\begin{equation}
\left(\frac{d x}{d\tau}\right)^2 = \Omega_{\Lambda,0} x^4 - \Omega_{k,0} x^2 +
\Omega_{m,0} x + \Omega_R = W(x).
\end{equation}
The only new solutions appear, when one considers $\Omega_{\omega}$ big
enough, for $\Omega_{R}$ to be negative. Otherwise, the solutions is
equivalent to that of a model with radiation only. We note that:
\begin{equation}
    W(0)=\Omega_{R}<0 \label{zero}
\end{equation}
applies to all models, restricting the solutions.
 
The most general form of the solution, satisfying $x(0)=1$, reads
\begin{multline}
x = 1+\frac{\wp'(\tau)+\Omega_{\Lambda,0}}
    {2[\wp(\tau)-\tfrac{1}{12}(6\Omega_{\Lambda,0}-\Omega_{k,0})]^2-\tfrac12\Omega_{\Lambda,0}}+\\
    +\frac{\tfrac12(4\Omega_{\Lambda,0}-2\Omega_{k,0}+\Omega_{\text{m},0})[\wp(\tau)-\tfrac{1}{12}(6\Omega_{\Lambda,0}-\Omega_{k,0})]}
    {2[\wp(\tau)-\tfrac{1}{12}(6\Omega_{\Lambda,0}-\Omega_{k,0})]^2-\tfrac12\Omega_{\Lambda,0}},
\end{multline}
with the invariants
\begin{align}
g_2 &= \tfrac{1}{12}\Omega_{k,0}^2+\Omega_{\Lambda,0}\Omega_{R}, \notag\\
g_3 &= \tfrac{1}{216}\Omega_{k,0}^3 - \tfrac{1}{16}\Omega_{\Lambda,0}\Omega_{\text{m},0}^2 - \tfrac16\Omega_{\Lambda,0}\Omega_{k,0}\Omega_{R}.
\end{align}
and using it, we can also obtain an explicit formula connecting the conformal 
and cosmological times.

\begin{figure}
\begin{center}
\includegraphics[width=0.9\textwidth]{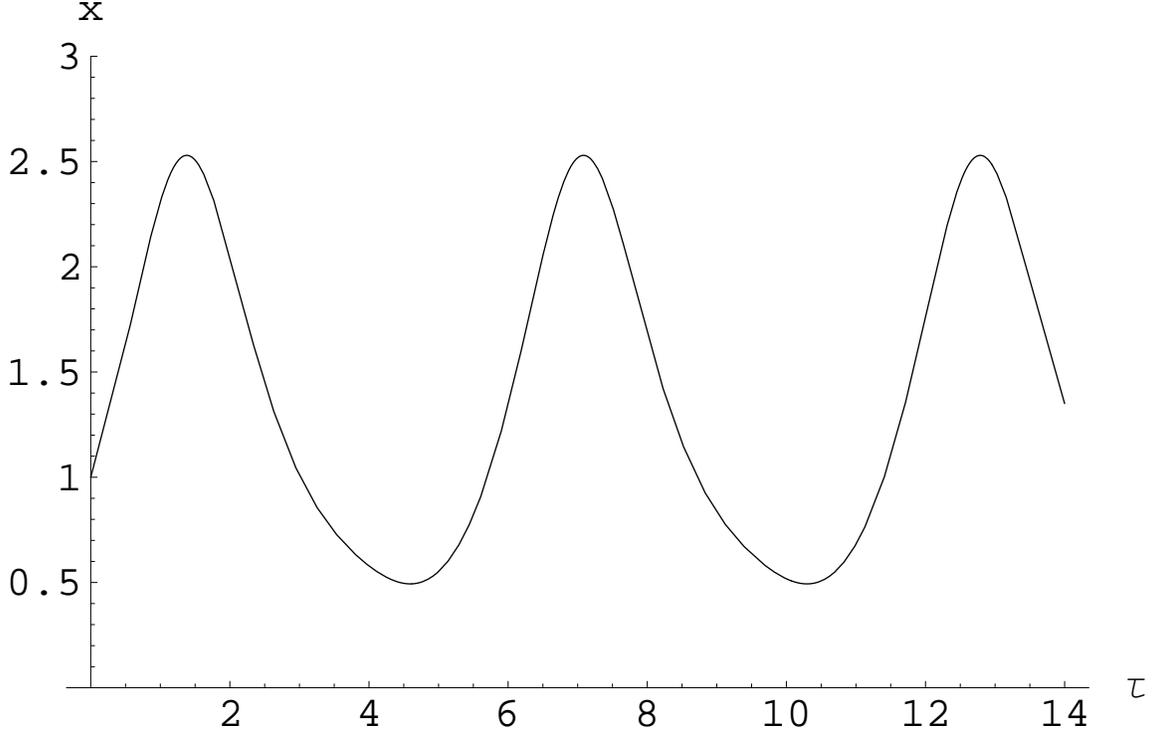}
\end{center}
\caption{ A generic oscillating solution, which in the considered case is 
never singular. Scaling from $\tau$ to $u$ only results in stretching the 
time axis.}
\label{rot_period}
\end{figure}

First, we need to investigate the singular points of our solution. As
$x(\tau)$ is an elliptic function of order two, there are exactly two poles
(provided that $\Omega_{\Lambda,0}\ne 0$). To specify them exactly, we can
use the $\wp$ function, together with its derivative
\begin{align}
    \wp(\tau_1) &= +\tfrac12\sqrt{\Omega_{\Lambda,0}} + \tfrac12\Omega_{\Lambda,0} - \tfrac{1}{12}\Omega_{k,0}, &
    \wp'(\tau_1) &= +\tfrac14\sqrt{\Omega_{\Lambda,0}} + \Omega_{\Lambda,0}, \notag\\
    \wp(\tau_2) &= -\tfrac12\sqrt{\Omega_{\Lambda,0}} + \tfrac12\Omega_{\Lambda,0} - \tfrac{1}{12}\Omega_{k,0}, &
    \wp'(\tau_2) &= -\tfrac14\sqrt{\Omega_{\Lambda,0}} + \Omega_{\Lambda,0}.
\end{align}
Relation $d u = x d\tau$ then becomes
\begin{equation}
    u = \left\{1 + \frac{1}{\sqrt{\Omega_{\Lambda,0}}}[\zeta(\tau_1)-\zeta(\tau_2)]\right\}\tau +
        \frac{1}{\sqrt{\Omega_{\Lambda,0}}}\ln\left[\frac{\sigma(\tau-\tau_1)\sigma(\tau_2)}
            {\sigma(\tau-\tau_2)\sigma(\tau_1)}\right],
\label{tautou}
\end{equation}
where the constant of integration was choosen so that $u(\tau=0)=0$, and
$\zeta$ and $\sigma$ are the appropriate Weierstrass functions.
 
This formula remains true in the degenerate cases as well, with $\zeta$ and
$\sigma$ also simplified. In the case of trigonometric/hyperbolic solutions
we have
\begin{align}
    \zeta(\tau) &= \tfrac12\left[ g\tau + \sqrt{6g} \cot(\sqrt{\tfrac32g}\,\tau)\right], \notag\\
    \sigma(\tau) &= \sqrt{\frac{2}{3g}}\,\mathrm{e}^{\tfrac14g\tau^2}\sin(\sqrt{\tfrac32g}\,\tau),
\end{align}
with $\sqrt3g=\pm\sqrt{g_2}$, and for initial conditions chosen as for the
general solution.

The two following sections consist of the case by case classification
analogous to that of \cite{Dabrowski:2004hx} with the quantities
$\sigma_i,w_-,x_{+/-/0}$ defined therein.

\subsection*{1. $\Omega_{\Lambda,0}>0$}

\begin{figure}
\begin{center}
\includegraphics[width=0.9\textwidth]{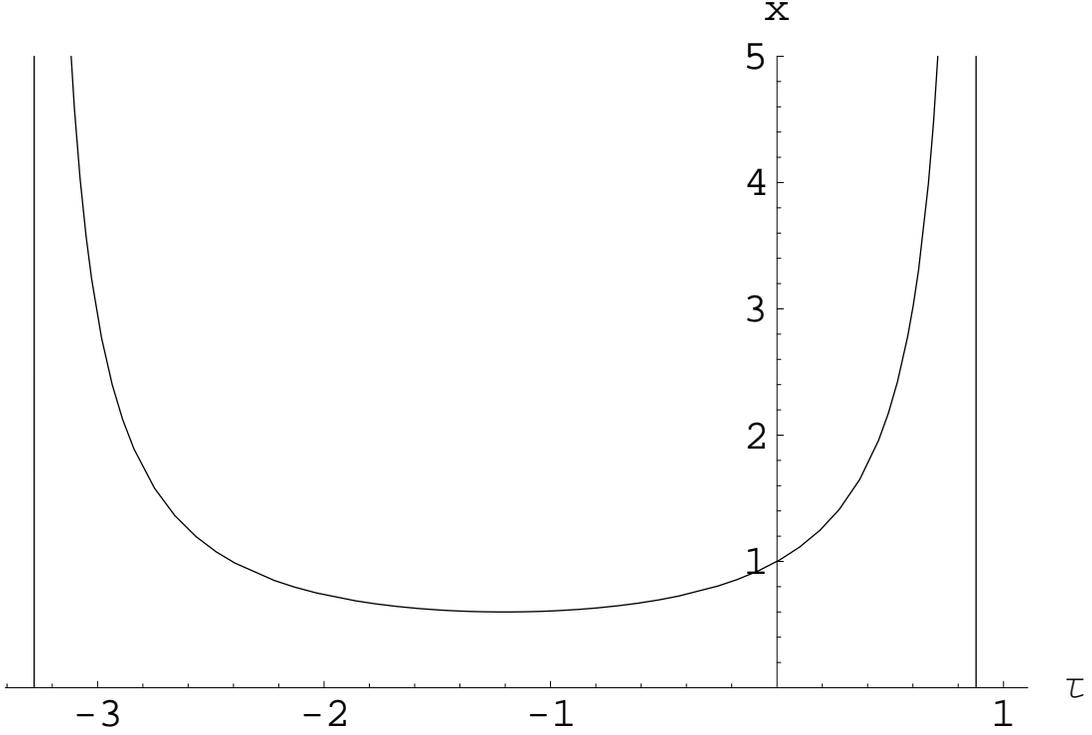}
\end{center}
\caption{ A typical bounce in finite time $\tau$,
corresponding to $u$ changing from $-\infty$ to $\infty$.}
\label{rot_fbounce}
\end{figure}

\subsection*{1.1. Case $\sigma_6 \ne 0$}
 
Analyzing the behavior of $\sigma_6$, with the possible values of
$\Omega_{R}$ and $\Omega_{\Lambda,0}$,
we obtain the first important restriction:
\begin{equation}
    \Omega_{k,0} \leqslant 0 \Rightarrow \sigma_6 < 0. \label{curv1}
\end{equation}
For positive curvature, any sign is possible.
 
As follows from the general considerations, we have two possibilities:
$\sigma_6 > 0$, or $\sigma_6 < 0$. However, the former is limited to the
case of four real roots. This happens because we have:
$w_- = (2\Omega_{k,0} - \sqrt{\Omega_{k,0}^2+12\Omega_{R}\Omega_{\Lambda,0}})/3$,
and by (\ref{curv1}) $\Omega_{k,0} > 0$ while
$\Omega_{R}\Omega_{\Lambda,0} < 0$, causing $w_-$ to be always positive,
and making the resolvent cubic admit three solutions.

\subsubsection*{1.1.1 Four complex roots}
 
This case is impossible, because $\Omega_{\Lambda,0}>0$ would force $w$ to
be positive everywhere.

\subsubsection*{1.1.2. Four real roots}
 
There are two possibilities here. Either $e_1<0<e_2<e_3<e_4$, or
$e_1<e_2<e_3<0<e_4$.
The former admits both the solutions $x_0$ and $x_+$, the latter only $x_+$.
However, none of them passes through zero, allowing for the particularly
interesting non-singular, periodic behavior of $x_0$. A typical oscillatory
solution is presented in Fig.~\ref{rot_period}, and the bounce in 
Fig.~\ref{rot_fbounce}. Although the latter has a finite time of evolution in
$\tau$, this interval is stretched into the whole real line when in the
variable $u$. This happens because the right hand side in (\ref{tautou})
is singular.

\subsection*{1.2. $\sigma_6=0$ -- one double root}

As a consequence of (\ref{curv1}), this, and obviously all the following
cases, require \emph{positive curvature}. Also, as in the general
classification, we must have:
\begin{equation}
    \Omega_{k,0}^2 + 12\Omega_{R}\Omega_{\Lambda,0} > 0.
\end{equation}
As $\sigma_6=0$, is effectively an equation of degree two in
$\Omega_{\Lambda,0}$ it instantly provides new restrictions. First, for it
to have real solutions in $\Omega_{\Lambda,0}$, we must have
\begin{equation}
    9\Omega_{m,0}^2 + 32\Omega_{k,0}\Omega_{R} \geqslant 0. \label{delta}
\end{equation}
and if at least one of the solutions is to be positive, we also need
\begin{equation}
    \frac{16\Omega_{R}\Omega_{k,0}}{\Omega_{\text{m},0}^2} \in \left[-3(3+\sqrt3);-3(3-\sqrt3)\right], \notag
\end{equation}
finally yielding:
\begin{equation}
    \frac{16\Omega_{R}\Omega_{k,0}}{\Omega_{\text{m},0}^2} \in \left[-\tfrac92;-3(3-\sqrt3)\right].
\end{equation}
 
This also allows us to see that the sign of $\sigma_5$, which is the same
as that of the expression
$\Omega_{\Lambda,0}-2k^3/(9\Omega_{\text{m},0}^2+8\Omega_{k,0}\Omega_{R})$
in this case, can be both plus and minus.
The biggest restriction comes from the inequality (\ref{zero}), which makes
case 1.2.1 impossible. As for 1.2.2, all three subcases might happen. If the
double root is the smallest, then we only have the $x_{-1}$ solution (bounce).
If it lies between the other roots, we have both the static, stable solution
$x_0$, and $x_+$ (bounce).
Lastly, if the double root is the biggest, we have both asymptotic branches
$x_+$ and $x_{-1}$, together with the unstable, static $x_0$.

\begin{figure}
\begin{center}
\includegraphics[width=0.9\textwidth]{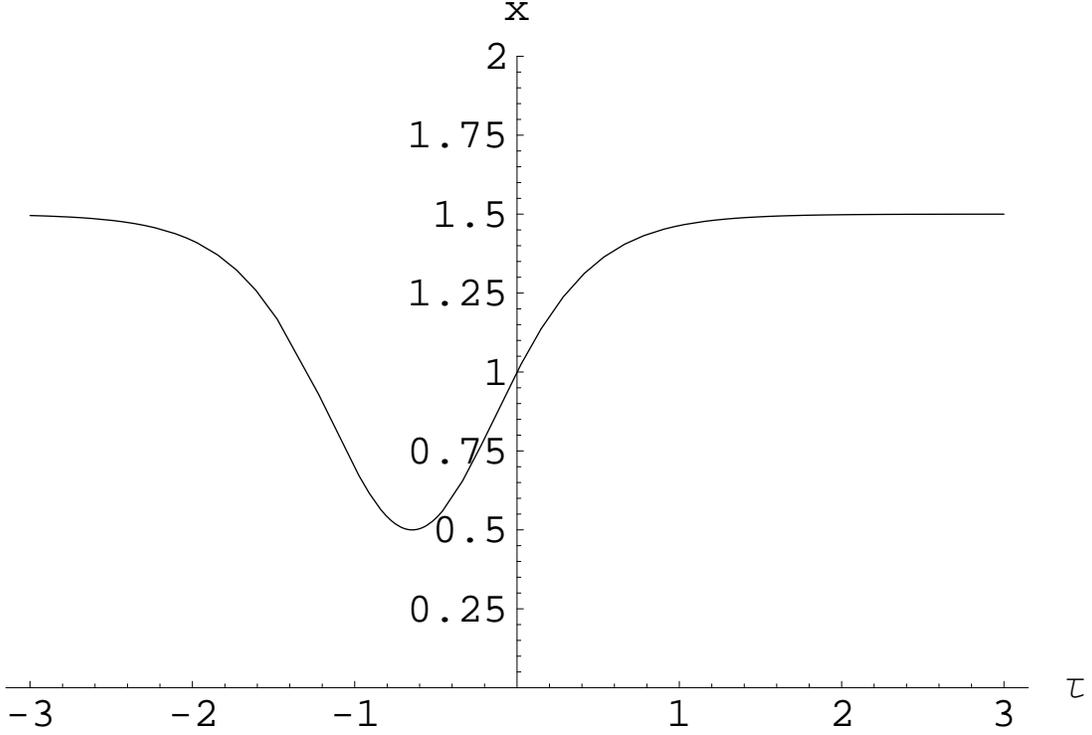}
\end{center}
\caption{The quasi-static solution, which stays
almost at the same value of the scale factor for most of the time
(both $\tau$ and $u$), and undergoes only one short dip.}
\label{rot_asympt1}
\end{figure}

\begin{figure}
\begin{center}
\includegraphics[width=0.9\textwidth]{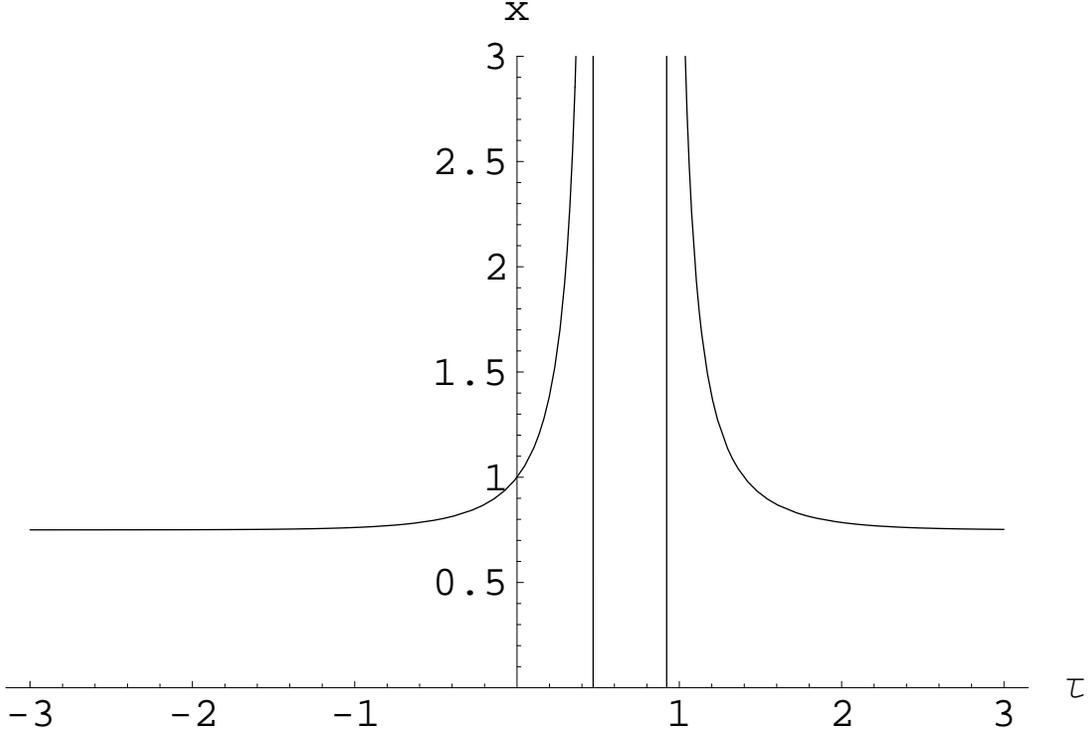}
\end{center}
\caption{Only the increasing solution in this
figure is physical, it tends to the static solution at minus infinity
(both $\tau$ and $u$) and the expansion to infinity takes infinite
time ($u$).}
\label{rot_asympt2}
\end{figure}

The bouncing solutions are essentially the same as the previous ones 
(Fig.~\ref{rot_fbounce}), also with time stretching. As for the asymptotic 
solutions, we either have a quasi-static evolution $x_+$ with one minimum, 
depicted in Fig.~\ref{rot_asympt1}; or a monotonic expansion/contraction 
$x_{-1}$, presented in Fig.~\ref{rot_asympt2}. The former has an infinite 
evolution time in both $\tau$ and $u$, and the latter reaches infinite $x$ for 
finite $\tau$ but infinite $u$.

\subsection*{1.3. $\sigma_6=\sigma_5=0$, $\sigma_4\ne 0$ -- two double roots}

Further simplifications require
\begin{align}
    \Omega_{\Lambda,0} &= \frac{\Omega_{k,0}^2}{12\Omega_{R}} \notag\\
    \Omega_{\text{m},0} &= \sqrt{-\tfrac{32}{9}\Omega_{k,0}\Omega_{R}}. \label{cond_triple}
\end{align}
However, such a value of $\Omega_{\Lambda,0}$ makes $\sigma_4=0$, making this 
case impossible, and bringing us to the next one.

\subsection*{1.4. $\sigma_6=\sigma_5=\sigma_4=0$, $\sigma_3\ne 0$ -- one triple root}

The so far excluded possibility of
$12\Omega_{R}\Omega_{\Lambda,0} + \Omega_{k,0}^2 = 0$,
holds here, and substituting (\ref{cond_triple}) into $\sigma_3$, we see that
further simplification is not possible, as $\sigma_3$ is proportional to
$\Omega_{R}^3 \ne 0$. Following the general classification 1.4, we can see
that the multiple root must be negative, because of (\ref{zero}) the simple
root must be positive, and thus only  the bounce solution can be physical.
Again, Fig.~\ref{rot_fbounce} applies.
 
\subsection*{2. $\Omega_{\Lambda,0}<0$}

\subsection*{2.1. $\sigma_6\ne 0$ -- simple roots}
 
As before we obtain a restriction on $\sigma_6$ 
\begin{equation}
    \Omega_{k,0} < 0 \Rightarrow \sigma_6 > 0 \label{curv2},
\end{equation}
and similarly to before, for non-negative curvature any sign is possible.
Analyzing the subcases of $\sigma_6>0$, we can see that one of the
restrictions becomes an identity now. Namely:
\begin{equation}
    \Omega_{k,0}^2 + 12\Omega_{R}\Omega_{\Lambda,0} >0
\end{equation}
is always true. Furthermore, for $k\leqslant 0$, we can only have $w_-<0$,
but for $k>0$ both signs are possible. In consequence, we can have the
subcase 2.1.2, which is all the more interesting thanks to (\ref{zero}),
which ensures, that there are oscillations with no singularity; and 2.1.3
with two possible regions of oscillation, also without singularity. They are
qualitatively the same as the $\Omega_{\Lambda,0}>0$ oscillations depicted
in Fig.~\ref{rot_period}.

\subsection*{2.2. $\sigma_6=0$ -- one double root}

Because of (\ref{curv2}), from now on we must have
\emph{non-negative curvature}. Obviously, (\ref{delta}) must still hold,
but the intervals of $\Omega_{R}$ are now ``reversed'':
\begin{equation}
    \frac{16\Omega_{R}\Omega_{k,0}}{\Omega_{\text{m},0}^2} \in 
    \left(-\infty;-3(3+\sqrt3)\right]\cup \left[-3(3-\sqrt3);0\right), \notag
\end{equation}
finally giving:
\begin{equation}
    \frac{16\Omega_{R}\Omega_{k,0}}{\Omega_{\text{m},0}^2} \in 
    \left[-3(3-\sqrt3);0\right).
\end{equation}
Looking at $\sigma_5$ as a function of $\Omega_{\Lambda,0}$ with the above
restrictions, it is straightforward to check that it can only be negative,
thus reducing this case to 2.2.1 only, where only a static, stable solution
exists.

\subsection*{2.3. $\sigma_6=\sigma_5=0$, $\sigma_4\ne 0$ -- two double roots}

As $\sigma_5=0$ has only positive roots with all the current restrictions on
the parameters, this and all the further cases are impossible.

\end{document}